# Vector Beams with Parabolic and Elliptic Cross-Sections for Laser Material Processing Applications

Sergej Orlov[1], Vitalis Vosylius[1], Pavel Gotovski[1], Artūras Grabusovas[1], Justas Baltrukonis[1], Titas Gertus[1,2]

[1] *State research institute Center for Physical Sciences and Technology, Industrial Laboratory for Photonic Technologies, Sauletekio ave 3 LT-10222, Vilnius, Lithuania*
*E-mail: sergejus.orlovas@ftmc.lt*
[2] *UAB "Altechna R&D", Workshop of Photonics, Mokslininkų st. 6a, LT-08412, Vilnius, Lithuania*

Beam profile engineering, where a desired optical intensity distribution can be generated by an array of phase shifting (or amplitude changing) elements is a promising approach in laser material processing. For example, a spatial light modulator (SLM) is a dynamic diffractive optical element allowing for experimental implementations of controllable beam profile. Scalar Mathieu beams have elliptical intensity distribution perceivable as "optical knives" in the transverse plane and scalar Weber beams have a parabolic distribution, which enables us to call them "optical shovels". Here, we introduce vector versions of scalar Mathieu and Weber beams and use those vector beams as a basis to construct controllable on-axis phase and amplitude distributions with polarization control. Further, we generate individual components of optical "knife" and "shovel" beams experimentally using SLMs as a toy model and report on our achievements in the control over the beam shape, dimensions and polarization along the propagation axis.

DOI: 10.2961/jlmn.2018.03.0023

## 1. Introduction

Laser beam shaping is an important technique used in modern laser beam applications such as light sheet microscopy, microfabrication and photopolimerisation to name a few. In situations where the same pattern is needed over long propagation distance it is advantageous to use non-diffracting beams. One of them is Bessel beam [1], which exhibits a long focal line with high length to width ratio [2]. Due to this property it can be perceived as an optical needle [3]. Optical needle beams are advantageous to use in applications such as fabrication of long canals in bulk material, trapping many particles simultaneously [2,4,5].

In some cases the microfabrication process is sensitive to the polarization structure of the laser beam [6]. Nonhomogeneous polarizations like azimuthal or radial have been shown to affect the efficiency of laser drilling procedure [7,8]. The polarization control of optical needle could also increase the speed of the laser microprocessing of the materials and will be discussed in this publication as an additional degree of freedom in the use of vectorial beams.

In a similar fashion, additional degree of freedom in the transverse profiles with asymmetrical intensity distributions, which have comparable to Bessel beam properties, can be achieved after introducing the so-called Mathieu-Gaussian and parabolic-Gaussian (Weber-Gaussian) beams. [9-11].

Mathieu beams possess a rather complicated distribution of electric field (sometimes called an "optical knife") which also has some practical applications due to its asymmetrical cross-section [12,13]. Scalar Weber beams have a distinct parabolic cross-section, which enables us to call them "optical shovels" [10]. The spatial shape of such beam can be controlled via the so-called parabolicity parameter, which enables their usage in various applications, where the transverse intensity profile is crucial.

Thus, both families exhibit non-diffracting properties similar to Bessel beams, where a relatively long focal depth retains unchanging intensity distribution, which makes them a promising candidate in laser processing. Control of the focal line and engineering its transverse profile [14-16] is the next step towards engineering of focal lines.

Here, we introduce vector versions of those beams with controllable polarization and investigate numerically their spatial spectra. We use vector Mathieu beams [17] and vector Weber beams as a basis to construct controllable on-axis phase and amplitude distributions with polarization control. Further, we attempt to generate components of vector Mathieu beams experimentally using SLMs and report on our achievements in the control over the beam shape and dimensions along the propagation axis.

## 2. Non-diffracting beams

Nondiffracting beams can be introduced while searching for solutions of the scalar Helmholtz equation

$$\nabla^2 \psi + k^2 \psi = 0, \qquad (1)$$

which can be easily solved using the method of the separation of the variables

$$\psi(x, y, z, t) = R(x, y)Z(z, t), \qquad (2)$$

where $Z$ is the longitudinal and time dependent part and the function $R$ represents the part of the solution, which is transverse. Textbooks on mathematical physics name four (Cartesian, polar, elliptic and parabolic) different cylindrical coordinate systems, where this method leads to the expression of the electric field as an integral over plane waves, which are situated on the cone defined by the transverse component of the wave vector $k_\rho$:

$$E(x,y,z,t) = e^{i(k_z z - \omega t)} \int_0^{2\pi} A(\phi) e^{ik_\rho[x\cos(\phi)+y\sin(\phi)]} d\phi. \quad (3)$$

The spatial spectra $S(k_\rho,\phi)$ of this electromagnetic field can be to the angular function $A(\phi)$ as [18, 19]

$$S(k_\rho,\phi) = (2\pi)^{-1} A(\phi) \delta\left(k_\rho - \sqrt{k^2 - k_z^2}\right). \quad (4)$$

According to the Morse and Feshbach, see Ref. [19], a scalar solution of the Helmholtz equation can lead to a solution of the vector wave equation, if following operations are applied [19]

$$\vec{M} = \nabla \times [\vec{a}\psi(\vec{r},q)],$$
$$\vec{N} = \frac{1}{k} \nabla \times \vec{M}. \quad (5)$$

Here a constant vector $\vec{a}$ defines some internal symmetries of the vector beams. The first solution is the transverse magnetic mode of the vector Helmholtz equation and the second one represents a transverse electric mode. In an obvious manner, the dual electromagnetic field can be handily expressed using those two solutions as a basis functions with coefficients $a_n$ and $b_n$, leading to

$$\vec{E} = -\sum_n (a_n \vec{M}_n + b_n \vec{N}_n),$$
$$\vec{H} = -\frac{k}{i\omega\mu} \sum_n (a_n \vec{N}_n + b_n \vec{M}_n). \quad (6)$$

Moreover, it can be shown, that the spatial spectra of a single mode from Eq. (5) can be expressed through the spatial spectrum of the scalar beam via [18,19]

$$\vec{F}_M(k_\rho,\phi) = i(\vec{k} \times \vec{a}) \cdot S(k_\rho,\phi),$$
$$\vec{F}_N(k_\rho,\phi) = \frac{1}{k}(\vec{k} \times \vec{a}) \times \vec{k} \cdot S(k_\rho,\phi). \quad (7)$$

In our case, we will be using $\vec{a} = \vec{e}_z$ and Eqs. (7) can be written as

$$\vec{F}_M(k_\rho,\phi) = [ik_0 \sin(\phi)\vec{e}_x - ik_0 \cos(\phi)\vec{e}_y] \cdot S(k_\rho,\phi)$$
$$\vec{F}_N(k_\rho,\phi) = [-k_0 \cos(\phi)\cos(\theta)\vec{e}_x - k_0 \sin(\phi)\cos(\theta)\vec{e}_y \quad (8)$$
$$+ k_0 \sin^2(\theta)\vec{e}_z] \cdot S(k_\rho,\phi).$$

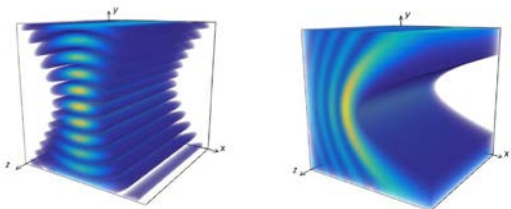

**Fig. 1** Intensity iso-surfaces of zeroth order even Mathieu with ellipticity parameter $q$=20 (left) and traveling wave Weber beams with $p$=4 (right).

### 3. Non-diffracting Mathieu beams

We define elliptic cylinder coordinates by the transformation

$$x + iy = \alpha \operatorname{acosh}(\xi + i\eta) \quad (9)$$

which introduces the so-called elliptical cylinder coordinates

$$x = \alpha \cosh(\xi)\cos(\eta),$$
$$y = \alpha \sinh(\xi)\sin(\eta), \quad (10)$$

here $\alpha$ is ellipticity parameter, $\eta,\xi$ are transverse elliptic coordinates. In these coordinates the three-dimensional Helmholtz equation separates into a longitudinal and transverse parts as given in Ref. [9,11]

$$\psi_m^e(\xi,\eta,z,t) = \psi_0 Je_m(\xi,q)ce_m(\eta,q)e^{i(\pm k_z z - \omega t)}; m \in 0,1,2...$$
$$\psi_m^o(\xi,\eta,z,t) = \psi_0 Jo_m(\xi,q)se_m(\eta,q)e^{i(\pm k_z z - \omega t)}; m \in 1,2,3...$$
$$\quad (11)$$

where, $Je_m$ is an even radial Mathieu function, $Jo_m$ - an odd radial Mathieu function, $ce_m$ is even angular Mathieu function and $se_m$ is an odd angular Mathieu function. A dimensionless parameter $q = \alpha^2 k_t^2/4$ and $k_t$ is a transverse wave vector component, $k_z$ is a longitudinal wave vector component and indices (*e*) and (*o*) correspond to the even and odd Mathieu beams of the order *m*. Helical Mathieu beams are defined as

$$\psi_m^h(\xi,\eta,z,t) = \psi_m^e(\xi,\eta,z,t) \pm i\psi_m^o(\xi,\eta,z,t). \quad (12)$$

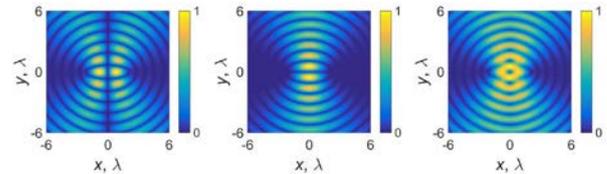

**Fig. 2** Transverse distribution of electric field of first order even, odd and helical (from the left to the right)) Mathieu beams.

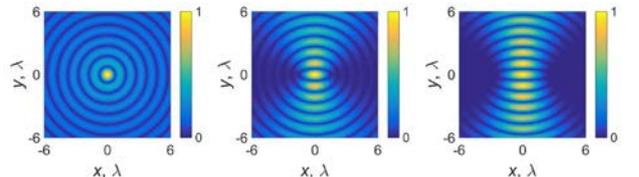

**Fig. 3** Transverse distribution of electric field of first order even Mathieu beam with different ellipticity parameter ($q$ = 0, 5, 20, from the left to the right).

In order to fully explain properties of nondiffracting pulses, full vectorial description must be used. Thus we vectorize scalar elliptic nondiffracting fields using [9,11]

$$M_x = \frac{\alpha}{h^2}\big[Je_m(\xi,q)ce'_m(\eta,q)\sinh(\xi)\cos(\eta)$$
$$+Je'_m(\xi,q)ce_m(\eta,q)\cosh(\xi)\sin(\eta)\big]e^{i(k_z z-\omega t)},$$
$$M_y = \frac{\alpha}{h^2}\big[Je_m(\xi,q)ce'_m(\eta,q)\cosh(\xi)\sin(\eta)$$
$$-Je'_m(\xi,q)ce_m(\eta,q)\sinh(\xi)\cos(\eta)\big]e^{i(k_z z-\omega t)},$$
$$M_z = 0.$$
$$\quad (13)$$

$$N_x = \frac{ik_z\alpha}{kh^2}\big[Je'_m(\xi,q)ce_m(\eta,q)\sinh(\xi)\cos(\eta)$$
$$-Je_m(\xi,q)ce'_m(\eta,q)\cosh(\xi)\sin(\eta)\big]e^{i(k_z z-\omega t)},$$
$$N_y = \frac{ik_z\alpha}{kh^2}\big[Je'_m(\xi,q)ce_m(\eta,q)\cosh(\xi)\sin(\eta)$$
$$+Je_m(\xi,q)ce'_m(\eta,q)\sinh(\xi)\cos(\eta)\big]e^{i(k_z z-\omega t)},$$

$$N_z = \frac{4q}{\alpha^2 k} Je_m(\xi,q) ce_m(\eta,q) e^{i(k_z z - \omega t)}. \tag{14}$$

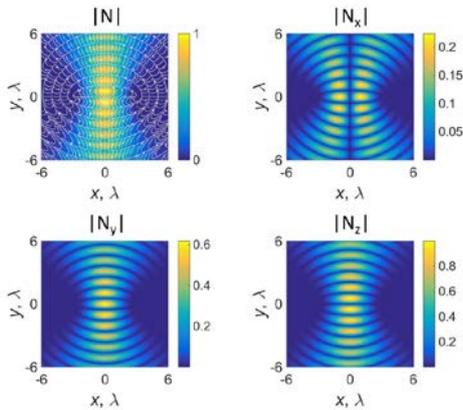

**Fig. 4** Transverse distribution of electric field of radially polarized first order even Mathieu beam and its components. White lines represent orientation of electric field. Ellipticity parameter $q = 20$.

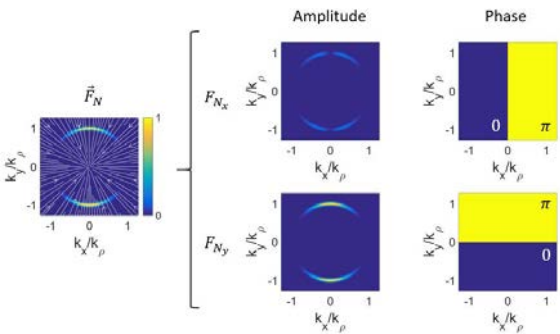

**Fig. 5** Angular spectrum of radially polarized zeroth order even Mathieu beam.

Transverse electric (TE) and transverse magnetic mode (TM) are obtained by choosing $\vec{a} = \vec{e}_z$. Fields in TE mode are azimuthally polarized and are described by a vector function $\vec{M}$ and fields in TM mode are radially polarized. A vector function $\vec{N}$ is used for their description. Linearly polarized Mathieu beams are weighted sum of TE and TM modes. Vector theory lets us describe Mathieu beams in optical systems with high numerical apertures as well when the incident field has a nonhomogeneous polarization.

Radially polarized Mathieu beam exhibit also a non-zero longitudinal component of the electric field, while azimuthally polarized fields are transverse only.

### 4. Non-diffracting Weber beams

We define parabolic cylinder coordinates by the transformation

$$\frac{\eta + i\xi}{\sqrt{2}} = \sqrt{x + iy}. \tag{15}$$

In these coordinates the three-dimensional Helmholtz equation separates into a longitudinal and transverse parts $\psi = U(\eta)V(\xi)Z(z)$. Longitudinal part has solution with dependence $\exp(ik_z z)$, and a transverse part

$$\frac{d^2 \Phi(\eta)}{d\eta^2} + (k_t^2 \eta^2 + 2k_t p)\Phi(\eta) = 0, \tag{16}$$

$$\frac{d^2 R(\xi)}{d\xi^2} + (k_t^2 \xi^2 - 2k_t p)R(\xi) = 0. \tag{17}$$

Here $k_t$ is a transverse vector and $p$ is a dimensionless parabolicity parameter. We change variables according to the rules $\sigma\xi \to v$ and $\sigma\eta \to u$, where $\sigma = \sqrt{2k_t}$, so Eqs. (16) and (17) are transformed into the canonical form of the parabolic cylinder differential equation

$$\frac{d^2 P}{dv^2} + (v^2/4 - p)P = 0 \tag{18}$$

Solutions to this differential equations are found by standard methods (e.g., Frobenius). So, even (*e*), odd (*o*) and traveling (*T*) nondiffracting parabolic beams are [11]:

$$U_e(\eta,\xi;p) = \frac{1}{\pi\sqrt{2}} |\Gamma_1|^2 P_e(\sigma\xi;p)P_e(\sigma\eta;-p),$$

$$U_o(\eta,\xi;p) = \frac{2}{\pi\sqrt{2}} |\Gamma_3|^2 P_o(\sigma\xi;p)P_o(\sigma\eta;-p). \tag{19}$$

$$TU^{\pm}(\eta,\xi;p) = U_e(\eta,\xi;p) \pm iU_o(\eta,\xi;p). \tag{20}$$

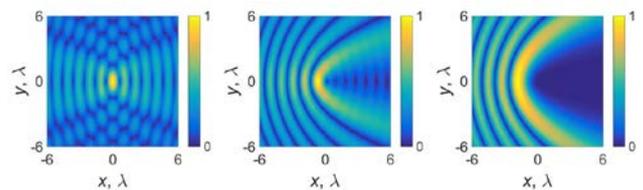

**Fig. 6** Transverse distribution of electric field of traveling wave Weber beam with different $p$ parameter ($p = 0, 1, 4$, from the left to the right).

where $\Gamma_1 = \Gamma(1/4 + ip/2)$ and $\Gamma_3 = \Gamma(3/4 + ip/2)$, here $\Gamma$ is the gamma function. We note that the odd and even type parabolic beams have not only positive but also negative $k_x$ and $k_y$ components in their spatial spectrum, therefore they do represent standing waves. On the other hand the travelling wave solution has either only positive $k_x$ or only positive $k_y$ (depending on the orientation of the coordinate system) In order to fully explain properties of nondiffracting pulses, full vectorial description must be used. Thus we vectorize scalar parabolic nondiffracting fields using $\vec{a} = \vec{e}_z$. An azimuthally polarized parabolic Weber beam is represented here by a vector field $\vec{M}$ and the radially – by a vector field $\vec{N}$:

$$\vec{M}(\vec{r},p) = \vec{e}_x \sigma e^{ik_z z} \frac{\xi U^{(0,1)}(\xi\sigma,\eta\sigma) + \eta U^{(1,0)}(\xi\sigma,\eta\sigma)}{\eta^2 + \xi^2}$$

$$+ \vec{e}_y \sigma e^{ik_z z} \frac{\eta U^{(0,1)}(\xi\sigma,\eta\sigma) - \xi U^{(1,0)}(\xi\sigma,\eta\sigma)}{\eta^2 + \xi^2}, \tag{21}$$

$$\vec{N}(\vec{r},p) = \vec{e}_x i\sigma e^{ik_z z} k_z \frac{\xi U^{(1,0)}(\xi\sigma,\eta\sigma) - \eta U^{(0,1)}(\xi\sigma,\eta\sigma)}{k(\eta^2 + \xi^2)}$$

$$+ \vec{e}_y i\sigma e^{ik_z z} k_z \frac{\xi U^{(0,1)}(\xi\sigma,\eta\sigma) + \eta U^{(1,0)}(\xi\sigma,\eta\sigma)}{k(\eta^2 + \xi^2)}$$

$$- \vec{e}_z \sigma^2 e^{ik_z z} \frac{U^{(0,2)}(\xi\sigma,\eta\sigma) + U^{(2,0)}(\xi\sigma,\eta\sigma)}{k(\eta^2 + \xi^2)}, \tag{22}$$

where $U^{(n,k)}(\xi\sigma,\eta\sigma) = \frac{\partial^{n+k} U(\xi\sigma,\eta\sigma)}{\partial \xi^n \partial \eta^k}$.

## 5. Engineering of optical focal lines

A longitudinal distribution of the beams intensity can be controlled by adding up vector Mathieu or Weber beams with different projections of the longitudinal wave vector. The axial intensity distribution and its spatial spectrum are associated via the Fourier transform

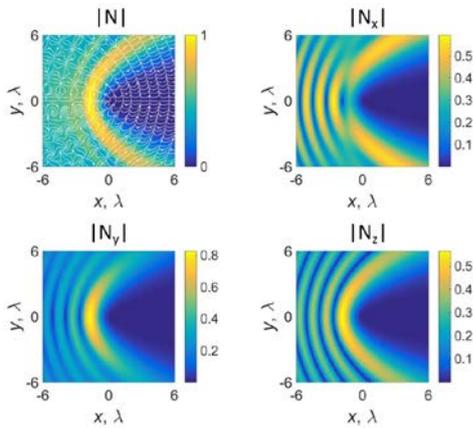

**Fig. 7** Transverse distribution of electric field of radially polarized traveling wave Weber beam and its components. White lines represent orientation of electric field. Parameter $p = 4$.

$$A(\beta) = \int f(z)\exp(-i\beta z)dz, \qquad (23)$$

where $\beta = k_z - k_{z0}$, and $k_{z0}$ is the projection of the carrier spatial frequency, $f(z)$ is a desired longitudinal intensity profile. Thus, the full control of optical beam can be obtained by experimentally realizing Eqs. (13,14) for Mathieu, Eqs. (21,22) for Weber together with Eq. (23).

For a radially polarized beam the spatial spectrum of a vector beam can be expressed as

$$\mathbf{\Psi}_N(k_x,k_y) = \int_{-\infty}^{\infty} A(\beta)\mathbf{F}_N(k_r,\varphi,q)d\beta. \qquad (24)$$

and for azimuthally it is

$$\mathbf{\Psi}_M(k_x,k_y) = \int_{-\infty}^{\infty} A(\beta)\mathbf{F}_M(k_r,\varphi,q)d\beta. \qquad (25)$$

As a toy model we choose two axial profile functions: 1) an axial distribution described by a boxcar (or a rectangular) function $f(z) = \Pi(Lz - L_1)$, where $L$ is its length, $L_1$ is the position and 2) an axial distribution described by a function $f(z) = \Pi(Lz - L_1) + \Pi(Lz - L_2)$, $L_1$ and $L_2$ are positions of two steps.

For the start, we choose the second axial distribution for the Mathieu based optical engineering. The spatial vector spectra of the optical beam is depicted in the Figure 8. We demonstrate a 3D distribution of an intensity isosurface of such beam for a situation with Mathieu beams (elliptical cross-section, see Section 3) and azimuthal polarization, Fig. 9. The cross-section of the beam in the middle of the first step is depicted there also, see Fig. 9.

For the Weber beams based beams, we choose the first axial distribution, described by the function $f(z) = \Pi(Lz - L_1)$. The spatial vector spectra of the optical beam is represented in the Fig. 10. We choose here the azimuthal polarization. We observe, that for an azimuthally polarized engineered profile, the structure of the vector spatial structure is different for $x$ and for $y$ component. Though they both contain a ring structure, the intensity of this structure changes differently. Thus, a generation of such complicated polarization structure will require not only phase and amplitude modulation but also independent modulation of $x$ and $y$ components.

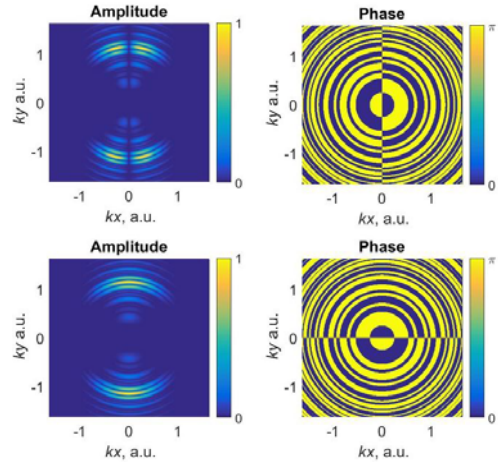

**Fig. 8** Angular spectrum of radially polarized zeroth order even Mathieu beam's $x$ (first row) and $y$ (second row) components with axial intensity distribution of two steps function.

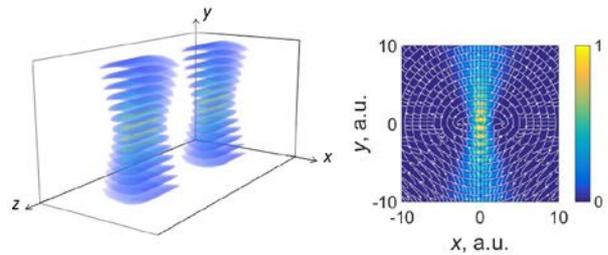

**Fig. 9** Intensity iso-surface (left) of a beam with axial intensity profile of two steps function and transverse intensity profile of a radially polarized zeroth order even Mathieu beam and (right) its transverse intensity distribution in a middle of the first step. White lines represents orientation of the electric field.

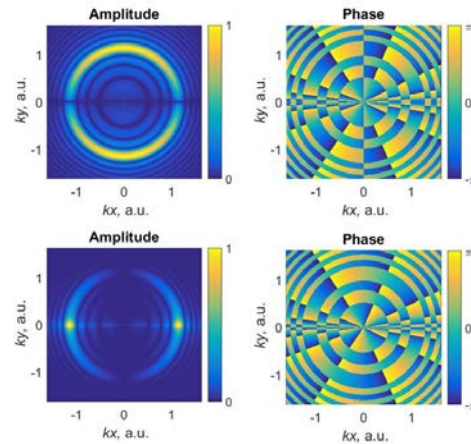

**Fig. 10** Angular spectrum of azimuthally polarized even Weber beam's $x$ (first row) and $y$ (second row) components with axial intensity distribution of a step function. The amplitude is in the first column and the phase of the components depicted in the second.

## 6. Experimental setup

The experimental setup for simulation of Mathieu and Weber beams (linear constituents of a vector field) is presented in the Figure 11.

Ideally one would use a single vectorial element, for an example, a transmissive geometrical phase (GP) element (something similar to the S-waveplate, see Ref. [20]) for generation of vector beams.

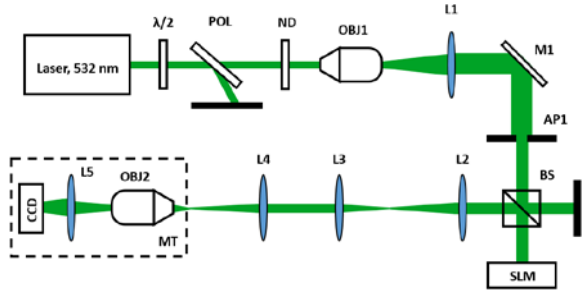

**Fig. 11** A sketch of the optical setup used in the experiment. CW laser, lenses (L1-L5), objectives (OBJ1 and OBJ2), a polarisator (POL), a mirror M1, aperture AP1, a beamsplitter BS, spatial light modulator (PLUTOVIS-006-A, HOLOEYE Photonics AG) and CCD camera.

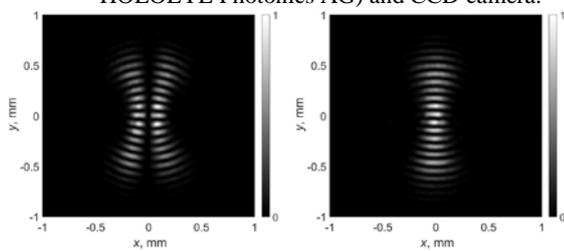

**Fig. 12** Intensity distribution of the experimentally obtained radially polarized zeroth order even Mathieu beam's $x$ (left) and $y$ (right) components with $q = 27$ in the transverse plane $(x,y)$.

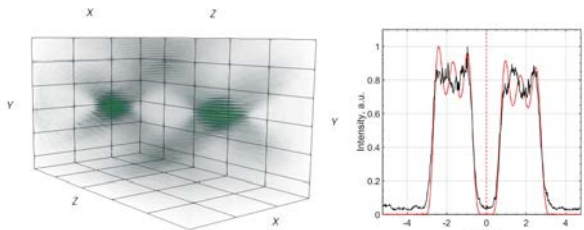

**Fig. 13** (a) A three-dimensional intensity iso-surface (left) of a beam with axial intensity profile of two steps function and transverse intensity profile of a radially polarized zeroth order even Mathieu beam's $y$ component and (right) its axial intensity profile (black) compared with numerical expectations (red). The red dotted line depicts the focal plane.

However, one can use as toy model of such GP element a phase-only SLM, where the phase mask encodes both amplitudes and phases of the spatial spectrum of an individual linear $x$- or $y$-component under the investigation. As a SLM requires the angle of incidence to be small, the beam splitter (BS) is used to enforce this and to propagate the diffracted beam further to the camera.
The procedure is as follows:
1. A linearly polarized and expanded light beam (at the wavelength of 532 nm) reaches the matrix of the SLM at angle of incidence of zero degrees.
2. It is reflected from the liquid crystal mask of the SLM, where the phase mask encodes both amplitude and phase spatial distribution. The 4f lens system (lenses L2 and L3) is used to transfer the spectral image from the SLM to a lens L4.
3. The lens L4 makes a Fourier transform of the beam generated on the SLM. The size of the liquid crystal mask of the SLM ($x_s$, $y_s$) and focal length of the Fourier lens $f_4$ determine spatial frequencies $f_x$ and $f_y$, which can be achieved in the setup.
4. Knowing the window of spatial frequencies ($f_x$, $f_y$), one can relate them to the maximal values of transverse wave vectors $k_x$, $k_y$ via $k = 2\pi f$. The pixel size $dx \times dy$ of the SLM determines the pixel size $dk_x \times dk_y$ of the spatial spectrum picture, which we obtain from Eqs. (24, 25).
5. The complex valued spatial spectrum has to be encoded to a phase-only picture, suitable for our SLM. The amplitude modulation is implemented here by using a checkerboard mask method, where groups of four phase-only pixels emulate a single pixel with arbitrarily chosen amplitude and phase [21]. In this method the amplitude A of the complex amplitude $A\exp(i\varphi)$ is encoded as the sum of two different phases for a 4x4 checkerboard: $\varphi_1 = \varphi - \arccos A$ and $\varphi_2 = \varphi + \arccos A$.
6. Thus, a desired beam profile is generated in the focal region of the lens L4. It is captured and recorded afterwards by the imaging system. A moving linear translation stage with mounted imaging system is moved along z axis.
7. While the stage moves, the transverse intensity data at different distances from focal plane is imaged onto the CCD matrix of the camera, this image is recorded and processed after-wards by a personal computer.
Our optical system achieves transverse magnification of approximately x27.4 and a longitudinal magnification of approximately x6.

Spatial spectrum of a vector beam is then expressed as a superposition of two linear components, which are then measured separately. We ensure the validity of our approach by adjusting amplitudes of the $x$ and $y$ components,

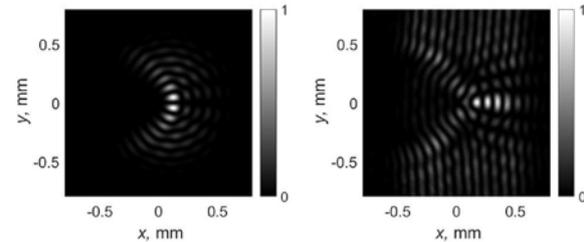

**Fig. 14** Intensity distribution of the experimentally obtained azimuthally polarized even Weber beam's $x$ (left) and $y$ (right) components with $p = 3$.

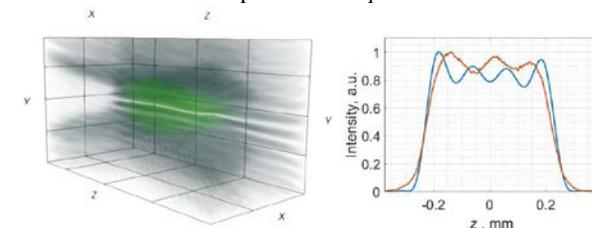

**Fig. 15** A three-dimensional intensity iso-surface (left) of a beam with axial intensity profile of a step function and transverse intensity profile of an azimuthally polarized even Weber beam's $x$ component and (right) its axial intensity profile (red) compared with numerical expectations (blue).

so they are proportional to the components of the actual vector beam.

### 7. Experimental results

For the sake of brevity and due to the lack of space, we restrict ourselves to a single polarization for each case of a different cross-section profile: elliptical beams will be presented only azimuthally polarized whereas only radially polarized Weber beams will be considered. Moreover, a single boxcar function will represent the axial profile of the Weber beams based optically engineered beams. A two step function $f(z) = \Pi(Lz - L_1) + \Pi(Lz - L_2)$ will be used as toy model for Mathieu beams based experiments.

First, we start with experimental verification of how vector Mathieu beams are generated. Experimental results are depicted in the Fig. 12. In general we observe a rather good agreement between our experimental results and numerical expectations, compare Fig. 8, 9. A three-dimensional depiction of the dominant *y*-component is presented in the Fig. 13. We note, that we observe on-axis, see Fig. 13 (right picture), some interference between two adjacent optical knives, which distorts the axial intensity profile near the focal plane (*z=0*), however those distortions are on the acceptable level.

Our preliminary analysis reveals, that this is happening due to the physical limitations of our experimental system – the dimensions of the liquid crystal matrix in the SLM are too small to properly represent the complex structure of the spatial spectrum.

Lastly, we demonstrate the outcomes of our experimental measurements for the case of azimuthally polarized Weber beams, with axial intensity distribution described by a function $f(z) = \Pi(Lz - L_1)$. Individual components are presented in the Fig. 14. Though we don't compare here those results directly with our numerical expectations, except briefly showing spatial spectra in Fig. 10, we can comment on a rather good agreement between our expectations from the numerical simulations and our experimental results.

A three-dimensional depiction of the dominant *x*-component is presented in the Fig. 15. We note, that we observe on-axis, see Fig. 15 (right picture) a rather good agreement between our experimental results and theoretical expectations, though some interference between two adjacent optical knives, which distorts the axial intensity profile near the focal plane (*z=0*), however those distortions are on the acceptable level.

## 8. Conclusions and outlook

In conclusion, we have presented a flexible theoretical technique, which enables us to create controlled axial profiles of vector optical "knives" (Mathieu based with elliptical crossection) and optical "shovels" (Weber beam based with parabolical crossection) with independent axial intensity profiles and polarization orientation. The method presented here may allow us to create various vector spatial structures with controllable on-axis polarization and axial intensity profile which might be applicable for possible specific microfabrication tasks or optical tweezing set-ups due to their elliptically or parabolically shaped transverse profiles.

A phase only spatial light modulator was successfully employed here as a toy model of an actual optical element in order to probe propagation of individual components of a radially polarized "optical knife" and azimuthally polarized "optical shovel" with two step axial profile.

Of course, implementation of this approach into high power laser systems will require us to move away from the conventional spatial light modulators, as they cannot sustain high laser powers (with some costly exceptions), however the spatial light modulator can be a versatile toy model for a geometrical phase element, which are known for their sustainability to high powers [20].

Nevertheless, we can report already on successful production of very first geometrical phase elements, designed with a know-how described in this report und suitable for simultaneous generation of both linearly polarized constituents of vector Mathieu and Weber beams based optical beams with controllable profiles and polarizations. Those results will be presented elsewhere.

**Acknowledgments**

This research is funded by the European Social Fund according to the activity 'Improvement of researchers' qualification by implementing world-class R&D projects' of Measure No. 09.3.3-LMT-K-712